\documentclass[twocolumn,useamsfonts]{pasj01}

\usepackage{mathrsfs} 
\usepackage{pifont} 
\usepackage{bm}
\usepackage[x11names,table]{xcolor}
\usepackage{fancybox}
\usepackage{url}
\usepackage{enumitem}
\usepackage{graphicx}

\newcommand{\diff}{{\mathrm d}}

\Received{$\langle$reception date$\rangle$}
\Accepted{$\langle$acception date$\rangle$}
\Published{$\langle$publication date$\rangle$}

\begin{document}

\title{Fortran interface layer of the framework for developing particle simulator FDPS}

\author{Daisuke \textsc{Namekata}\altaffilmark{1}}
\email{daisuke.namekata@riken.jp}
\altaffiltext{1}{RIKEN Advanced Institute for Computational Science, 7-1-26 Minatojima-minami-machi, Chuo-ku, Kobe, Hyogo 650-0047, Japan}

\author{Masaki \textsc{Iwasawa}\altaffilmark{1}}
\email{masaki.iwasawa@riken.jp}

\author{Keigo \textsc{Nitadori}\altaffilmark{1}}
\email{keigo@riken.jp}

\author{Ataru \textsc{Tanikawa}\altaffilmark{1,2}}
\email{tanikawa@ea.c.u-tokyo.ac.jp}
\altaffiltext{2}{Department of Earth and Astronomy, College of Arts and Science, The University of Tokyo, 3-8-1 Komaba, Meguro-ku, Tokyo 153-8902, Japan}

\author{Takayuki \textsc{Muranushi}\altaffilmark{1}}
\email{takayuki.muranushi@riken.jp}

\author{Long \textsc{Wang}\altaffilmark{1,3}}
\email{long.wang@riken.jp}
\altaffiltext{3}{Argelander Institut F{\" u}r Astronomie, Auf Dem H{\" u}gel 71, 53121 Bonn, Germany}

\author{Natsuki \textsc{Hosono}\altaffilmark{1,4}}
\email{natsuki.hosono@riken.jp} 
\altaffiltext{4}{Yokohama Institute for Earth Sciences, Japan Agency for Marine-Earth Science and Technology, 3173-25, Showa-machi, Kanazawa-ku, Yokohama-city, Kanagawa, 236-0001, Japan}

\author{Kentaro \textsc{Nomura}\altaffilmark{1}}
\email{kentaro.nomura@riken.jp}

\author{Junichiro \textsc{Makino}\altaffilmark{1,5,6}}
\email{makino@mail.jmlab.jp}
\altaffiltext{5}{Department of Planetology, Graduate School of Science, Kobe University, 1-1 Rokkodai-cho, Nada-ku, Kobe 657-8501, Hyogo, Japan}
\altaffiltext{6}{Earth-Life Science Institute, Tokyo Institute of Technology, 2-12-1 Ookayama, Meguro-ku, Tokyo 152-8551, Japan}


\maketitle

\begin{abstract}
Numerical simulations based on particle methods have been widely used in various fields including astrophysics. To date, simulation softwares have been developed by individual researchers or research groups in each field, with a huge amount of time and effort, even though numerical algorithms used are very similar. To improve the situation, we have developed a framework, called FDPS, which enables researchers to easily develop massively parallel particle simulation codes for arbitrary particle methods. Until version 3.0, FDPS have provided API only for C++ programing language. This limitation comes from the fact that FDPS is developed using the template feature in C++, which is essential to support arbitrary data types of particle. However, there are many researchers who use Fortran to develop their codes. Thus, the previous versions of FDPS require such people to invest much time to learn C++. This is inefficient. To cope with this problem, we newly developed a Fortran interface layer in FDPS, which provides API for Fortran. In order to support arbitrary data types of particle in Fortran, we design the Fortran interface layer as follows. Based on a given derived data type in Fortran representing particle, a \textsc{Python} script provided by us automatically generates a library that manipulates the C++ core part of FDPS. This library is seen as a Fortran module providing API of FDPS from the Fortran side and uses C programs internally to interoperate Fortran with C++. In this way, we have overcome several technical issues when emulating `template' in Fortran. By using the Fortran interface, users can develop all parts of their codes in Fortran. We show that the overhead of the Fortran interface part is sufficiently small and a code written in Fortran shows a performance practically identical to the one written in C++.
\end{abstract}

\section{Introduction}
\label{sec:introduction}
Numerical simulations based on particle methods have been widely used in many fields of science and engineering. For example, in astrophysics, gravitational $N$-body and smoothed particle hydrodynamics (SPH) simulations are commonly used to study dynamics of celestial bodies. In the context of engineering and biology, moving particle semi-implicit (MPS) method, molecular dynamics (MD), power dynamics, and distinct element method (DEM) simulations are utilized for a variety of purposes such as disaster prevention and drug discovery. The reason of frequent use of particle-based methods may be attributed to the fact that various kinds of physical systems can be well modeled by collections of particles and particle-based methods have several advantages (e.g. adaptivity in space and time, Galilean invariance). The numbers of particles used in simulations go on increasing because there has been a rise in the need for refining results drawn from numerical simulations. For instance, cosmological $N$-body simulations can be used to study the evolution of rare objects such as quasars and active galactic nuclei. In order to make a prediction for such objects with a good statistical accuracy, we need to simulate large volume of the universe \citep{ishiyama15:_the_nu2_gc_simulations}. Large-scale simulations are also required in MD simulations of vapor-to-liquid nucleation for a direct comparison of predicted nucleation rates with those measured by laboratory experiments \citep{diemand13:_large_scale_molecular}. This trend requires researchers to develop simulation codes that run efficiently on modern distributed-memory parallel supercomputers.

However, developing such a code is quite difficult, time-consuming task because in order to achieve high efficiency we need to implement efficient algorithms such as the Barnes-Hut tree algorithm \citep{barnes_hut86:_a_hierarchical} as well as some load balancing mechanism into the code. To date, simulation codes have been developed by individual researchers or research groups in each field of science, with spending a huge amount of time and effort, even though numerical algorithms used are very similar to each other. Besides, in many cases, simulation codes are being developed for each specific application of particle methods, making it difficult for researchers to try a new method or another method. These are not efficient. 

In order to improve this situation, we have developed a framework FDPS (Framework for Developing Particle Simulators)\footnote{\url{https://github.com/FDPS/FDPS}} \citep{2015FDPS,iwasawa16:_implem_fdps} that enables researchers to easily develop codes for massively parallel particle simulations for arbitrary particle methods without taking care of parallelization of their codes. More specifically, FDPS provides a set of functions that (1) divide a computational domain into subdomains based on a given distribution of particles and assigns each subdomain to one MPI process, (2) exchange the information of particles among processes so that each process owns particles in its subdomain, (3) collect the information of particles necessary to perform interaction calculations in each process, and (4) perform interaction calculations using the Barnes-Hut tree algorithm for long-range force or a fast tree-based neighbor search for short-range force. All of these functions (hereafter called FDPS API [application programing interface]) are highly optimized and codes developed by using FDPS show good scalings for a large number of processes (up to $\sim 10^{5}$ processes; \citet{iwasawa16:_implem_fdps}). Thus, FDPS allows researchers to concentrate on their studies without spending time to the parallelization and complex optimization of the codes. FDPS has already been used to develop various applications (e.g. \cite{hosono16:_a_comparison_of_sph,hosono16:_the_giant_impact,hosono17:_unconv_of_very_large,michikoshi17a:_simulat_smallest,tanikawa17:_does_explo_nucl,iwasawa17:_pentacle_parallelized,tanikawa18:_tidal_double_detona,tanikawa18:_high_resolution_hydro}).

One issue of the previous versions of FDPS (version 2.0 or ealier) is that it requires researchers to develop codes in the C++ programing language. This limitation comes from the fact that FDPS is written in C++ to use the template feature in C++ to support arbitrary data types of particles. However, there are many supercomputer users who use mainly Fortran language to develop codes. In order for such people to use the functionalities of FDPS, they must first learn the C++ language with spending time. In addition, programs which they have written in Fortran in the past can no longer be used in newly-developed codes. To cope with this problem, we have redesigned FDPS for it to have a Fortran interface layer (hereinafter, we call it Fortran interface). This layer provides API for Fortran, i.e., a set of subroutines/functions to use the functionalities of FDPS from Fortran. Thus, in FDPS version 3.0 or later, researchers can develop their simulation codes in Fortran. In this paper, we present the basic design and implementation of Fortran interface and compare performance of a simple application written by using Fortran interface with that written in C++ by using FDPS. We will show that the overhead of Fortran interface is sufficiently small and the performances of both codes are almost identical.

This paper is organized as follows: In Section~\ref{sec:overview_of_fdps}, we briefly review the C++ core part of FDPS necessary to explain the design and implementation of Fortran interface, which are described in Section~\ref{sec:design_implementation}. Performance of application developed by using Fortran interface is shown in Section~\ref{sec:performance}. In Section~\ref{sec:example}, we present one example of practical application.  Finally, in section~\ref{sec:summary}, we summarize this study.

\section{Overview of FDPS}
\label{sec:overview_of_fdps}
In this section, we provide a brief overview of the previous versions (version 2.0 or earlier) of FDPS. In \S~\ref{subsec:what_fdps_is}, we describe what FDPS does and how FDPS looks like from users. In \S~\ref{subsec:fdps_impl}, we explain the implementation details of the previous versions of FDPS. 

\subsection{What FDPS does and its user interface design}
\label{subsec:what_fdps_is}
FDPS provides C++ library functions that perform tasks required for parallel particle simulations. In distributed-memory parallel supercomputers, particle simulations are generally performed in the following procedure.
\begin{enumerate}[label=(\arabic*)]
\item The computational domain is divided into subdomains based on a given distribution of particles and each subdomain is assigned to one MPI process. We call this task ``domain decomposition". 
\item Particles is exchanged among processes so that each process owns particles in its subdomain. We call this task ``particle exchange".
\item Each process collects the information necessary to perform interaction calculation for the particles belonging to this process.
\item Each process performs interaction calculation.
\item The information of particles is updated using the result of the interaction calculation in each process.
\end{enumerate}
Among the procedures above, procedures (1)-(3) are specific for parallel computation and the implementation of them into a code is a daunting task. As described in \S~\ref{sec:introduction}, FDPS provides APIs, a suite of functions, that perform these procedures efficiently. Therefore, by using FDPS, researchers can avoid the implementation of these complex procedures. In addition, FDPS also provide APIs that perform procedure (4), i.e. interaction calculation, with using fast algorithms for both long- and short-range forces. Thus, users of FDPS can perform these procedures just by calling FDPS APIs. The other parts of users' codes including procedure (5) are implemented by the users. These parts do not involve parallelization and thus users of FDPS do not need to consider the parallelization of the codes explicitly.

\begin{figure}[h]
\begin{center}
\includegraphics[width=8cm]{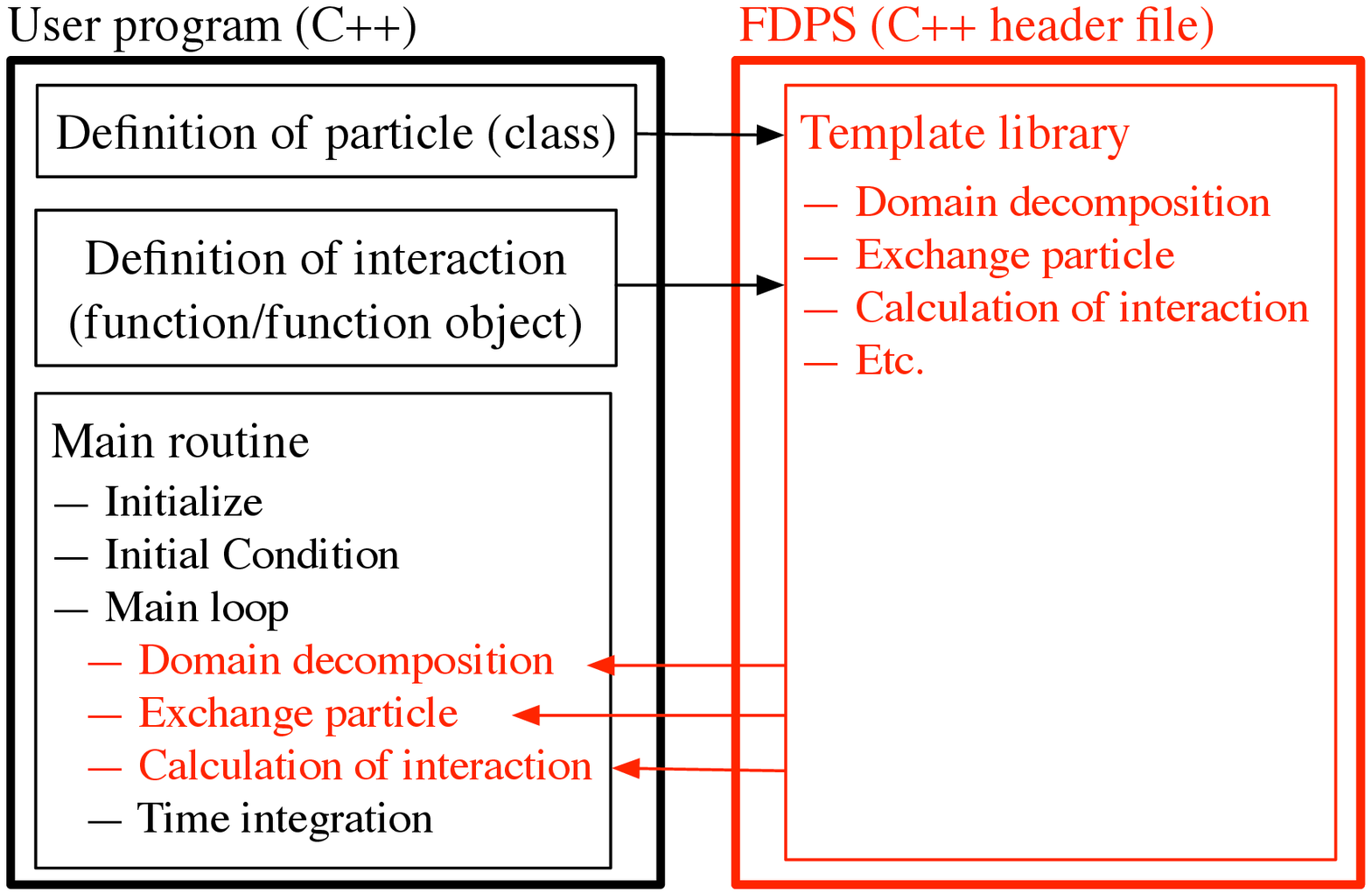}
\end{center}
\caption{A schematic illustration for the user interface design of the previous version of FDPS. Note that this figure is the same as Fig.1 in \citet{iwasawa16:_implem_fdps}.}
\label{fig:fdps_ui}  
\end{figure}

Figure~\ref{fig:fdps_ui} schematically illustrates typical structure of parallel particle simulation code developed by using FDPS and how a user uses FDPS. The user program first defines particle data set and interaction by using class and function in the C++ programing language, respectively. Thus, the user program must be written in C++. This is because, as shown in the right portion of Fig.~\ref{fig:fdps_ui}, FDPS is written in C++ by the reason explained in \S~\ref{subsec:fdps_impl}. FDPS provides APIs to perform domain decomposition, particle exchange, and calculation of interaction. They are defined as function templates in C++ (see \S~\ref{subsec:fdps_impl}). These function templates are combined with the definitions of particle and interaction in order to instantiate functions that perform domain decomposition, particle exchange, and calculation of interaction in the user program. The instantiated functions are used in the main loop of the user program as shown in the lower left portion of Fig.~\ref{fig:fdps_ui}. FDPS accepts arbitrary types of particle and interaction. Thus, we can use FDPS  to develop any kinds of particle simulation codes. 

In the next section, we will explain how the user interface design described here is realized by using features in the C++ programing language.

\subsection{Implementation details of FDPS}
\label{subsec:fdps_impl}
One of development goal of FDPS is to make it possible for FDPS to support arbitrary types of particle and interaction. In order to achieve this without a decrease in performance of simulation codes, FDPS is implemented by using the template feature in the C++ programing language. Roughly speaking, this feature allows functions and classes to be described by the use of general data types (for detail, see descriptions in the ISO C++ standard [ISO/IEC 14882:2014]). Functions and classes described by using the template feature are called function templates and class templates, respectively. All general data types used in a function template or class template are replaced by specific data types given through special arguments called template arguments at the compile-time of a program. Hence, there are no unknown data types at the compile-time, allowing compilers to aggressively optimize function templates. Thus, using the template feature, we can make a high performance library for general particle types. These are the reasons why we adopt the C++ programing language to develop FDPS.

In order to make it possible for FDPS to provide the APIs that performs domain decomposition, particle exchange, interaction calculation for arbitrary types of particle data and interaction, we adopt an internal structure for FDPS as described below. First, all functions in FDPS API are implemented as function templates to manipulate general particle types. As a result, users of FDPS must implement particle data as C++ classes and pass them to FDPS API through template arguments when using the API. Second, we require that particle classes defined by FDPS users must have some of public member functions that have specific names and specific functionalities. This is because FDPS, for example, must know which member variable represents the position of a particle to perform  domain decomposition and particle exchange. We solve this by requiring that a particle class has a public member function named \texttt{getPos} that returns the position of a particle. For similar reasons, there are other public member functions that a user-defined particle class should have. For a complete list, one can refer to the specification document of FDPS. Third, we require that particle-particle interactions must be implemented as functions or functors that have a specific interface. With this, we can implement FDPS without knowing the contents of interaction functions. This requirement does not restrict types of interactions; we can still implement arbitrary types of interaction functions as long as they have the specified interface.
 
The FDPS API is actually provided as the public member functions of two class templates \texttt{ParticleSystem} and \texttt{TreeForForce}, and one (non template) class \texttt{DomainInfo}. Hereinafter, they are called collectively FDPS classes. Each of them is used to make FDPS do a certain task. To be more precise, instances of FDPS classes \texttt{ParticleSystem}, \texttt{DomainInfo}, and \texttt{TreeForForce} are used, respectively, for particle exchange, domain decomposition, and interaction calculation.

\begin{figure}[h]
\begin{center}
\includegraphics[width=8cm]{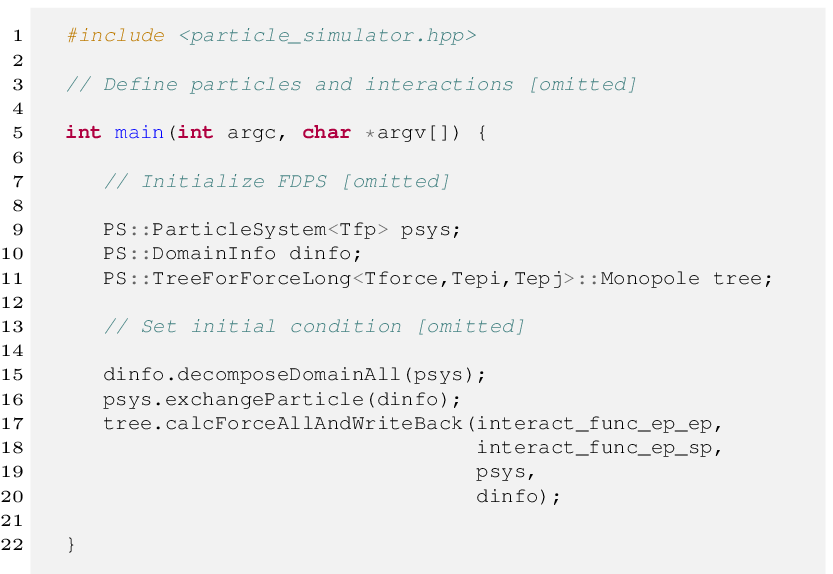}
\end{center}
\caption{An example of a C++ code which uses FDPS. }
\label{fig:src_cpp_smpl}  
\end{figure}

To illustrate the usage of FDPS API, we show in Figure~\ref{fig:src_cpp_smpl} an example of a C++ code that uses FDPS. At the first line of the example, the file \texttt{particle\_simulator.hpp} is included. This is the file where all FDPS classes are implemented. Hence, all user programs include this file. As explained above, in order to use the functionalities of FDPS, we first create instances of FDPS classes. This is done at lines 9--10 in the example, where ``\texttt{PS}" is the name space in which FDPS classes are defined and the words separated by commnas in the angle brackets associated with FDPS classes \texttt{ParticleSystem} and \texttt{TreeForForce} are template arguments. \texttt{ParticleSystem} class takes \texttt{FullParticle} (\texttt{FP}) class as a template argument, where \texttt{FP} class is a class that contains all information about a particle necessary to perform simulations as member variables. \texttt{Tfp} in the example is the user-defined \texttt{FP} class. In the example, \texttt{TreeForForceLong} class is used. It is a \texttt{TreeForForce} class specifically for long-range force and takes the following three particle classes as template arguments --- the first one is \texttt{Force} class which  contains the quantities of an $i$-particle used to store the results of the calculations of interactions as member variables; the second and third ones are \texttt{EssentialParticleI} (\texttt{EPI}) and \texttt{EssentialParticleJ} (\texttt{EPJ}) classes which contain the minimum sets of the quantities of $i$- and $j$-particles necessary to calculate interactions as member variables, where $i$-particle is a particle that receives a force and $j$-particle is a particle that exerts a force. In the example, \texttt{Tforce}, \texttt{Tepi}, \texttt{Tepj} are the user definitions of these three classes. Hereinafter, we call the four kinds of particle classes described above user-defined types\footnote{The user-defined types EPI and EPJ are introduced in order to reduce the requirements for the communication bandwidth and the memory bandwidth when performing interaction calculation.}. After creating the instances of FDPS classes, we can use the FDPS API by calling public member functions of these instances as in lines 15--20. 

More details about the previous version of FDPS can be found in \citet{iwasawa16:_implem_fdps}.

\section{Design, usage, and implementation of Fortran interface}
\label{sec:design_implementation}
In this section, we describe the details of our Fortran interface. In \S~\ref{subsec:ftn_if_ui}, we describe the user interface design of the Fortran interface. In \S~\ref{subsec:ftn_if_usage}, we explain the usage of Fortran interface with using a specific example of Fortran program. In \S~\ref{subsec:ftn_if_impl}, we describe the implementation details of Fortran interface. In the following, we call the part of FDPS corresponding to the previous version of FDPS the core part.

\subsection{User interface design of Fortran interface}
\label{subsec:ftn_if_ui}
\begin{figure}[h]
\begin{center}
\includegraphics[width=8cm]{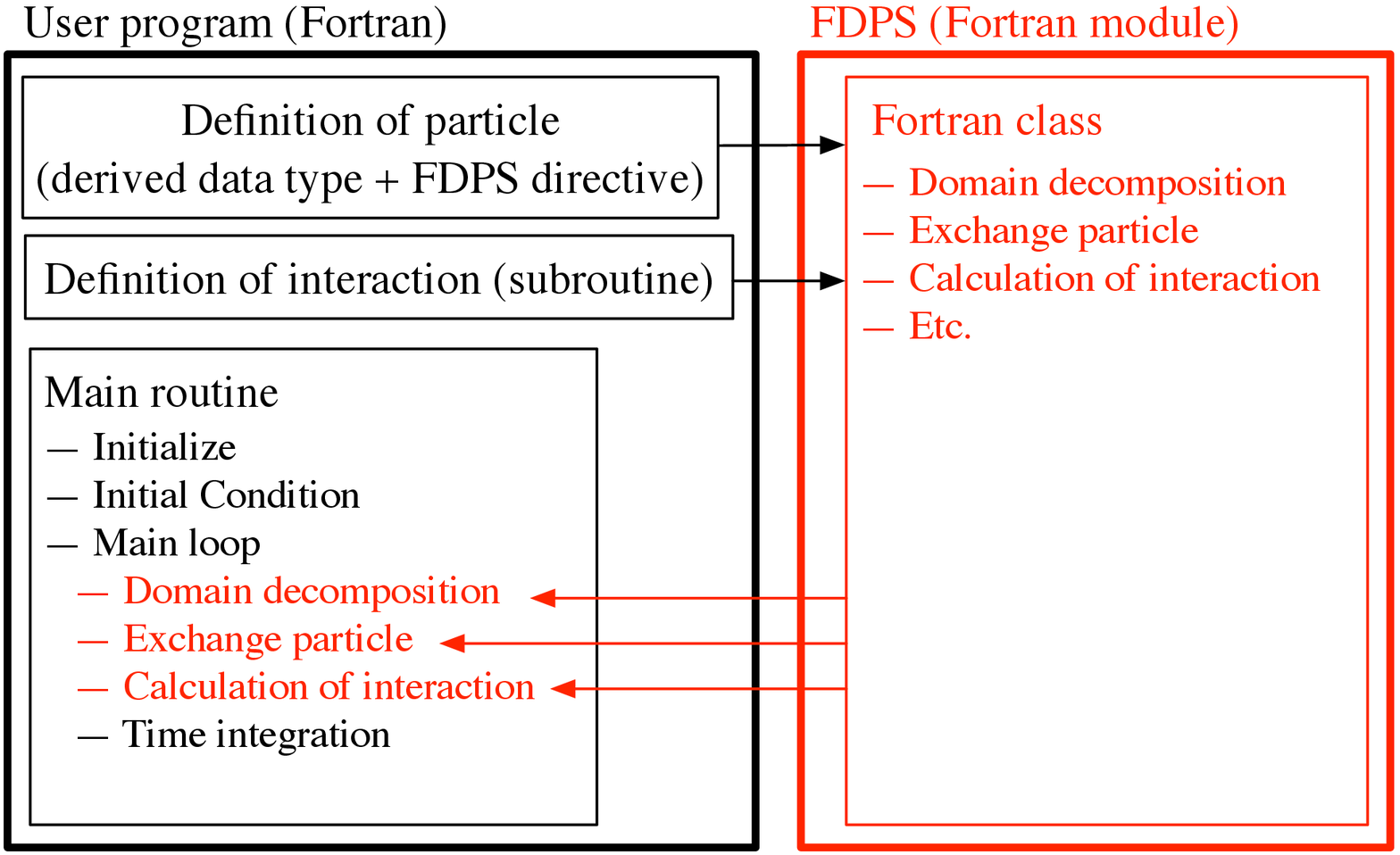}
\end{center}
\caption{A schematic illustration for the relation between user's Fortran program and FDPS Fortran interface.}
\label{fig:ftn_if_ui}  
\end{figure}

The Fortran interface to FDPS developed in this study wraps the core part of FDPS and hence has essentially the same user interface design as shown in Fig.~\ref{fig:ftn_if_ui}. The user program first defines particle and interaction by using derived data type and subroutine in Fortran 2003, respectively. Thus, the user program must be written in Fortran 2003. This is because, as will be explained in \S~\ref{subsec:ftn_if_impl}, the Fortran interface uses features in Fortran 2003 in order to make Fortran interoperate with C++. In the definition of particles, the user must describe complementary information about particle by using FDPS directives, which are Fortran comments having special format and are introduced by us (for details, see \S~\ref{subsec:ftn_if_usage}). This procedure corresponds to define specific public member functions in particle class when using FDPS from C++ (see \S~\ref{subsec:fdps_impl}). The Fortran interface to FDPS is written in a Fortran module as shown in the right portion of Fig.~\ref{fig:ftn_if_ui} and all of the FDPS API are defined as public member functions of a Fortran class for manipulating the core part of FDPS. Corresponding with the APIs in the core part of FDPS, there are APIs for domain decomposition, particle exchange, and calculation of interaction in Fortran interface. The user program calls these APIs in the main loop as shown the lower left portion of Fig.~\ref{fig:ftn_if_ui}.

The Fortran interface is also designed to accept arbitrary types of particle and interaction as with the core part of FDPS. In the core part of FDPS, this is realized by using the template feature in C++ as described in \S~\ref{subsec:fdps_impl}. Fortran does not have such feature. In the Fortran interface developed in this study, we realize it by an automatic generation of source programs of Fortran interface specifically for particle types given by users based on the information given through FDPS directives (the reason why this kind of mechanism is needed will be described in \S~\ref{subsec:ftn_if_impl}). The generation of source programs of Fortran interface is done by a \textsc{Python} script provided by us. The users of Fortran interface must run this script to use Fortran interface.

In the following subsections, we first explain the usage of Fortran interface in \S~\ref{subsec:ftn_if_usage} and then describe the implementation details of Fortran interface in \S~\ref{subsec:ftn_if_impl}.

\subsection{Usage of Fortran interface}
\label{subsec:ftn_if_usage}
In this section, we explain using a sample code how a user can write an application program using Fortran interface. The basic procedures to develop codes using Fortran interface are as follows.
\begin{enumerate}[label=\Roman*.]
\item Define a particle using a Fortran derived data type. 
\item Generate Fortran interface programs.
\item Define an interaction using a Fortran subroutine. 
\item Develop the main part of a code using the FDPS API given through Fortran interface. 
\end{enumerate}
Thus, the development procedures are similar to the case in which we develop codes in C++ using FDPS (see \S~\ref{sec:overview_of_fdps}). The only difference is the existence of Procedure II. There is no difficulty in this procedure because a user only have to execute the auto-generation script provided by us as stated earlier.

The sample code used is a gravitational $N$-body simulation code and is essentially same as the one shown in \S~2.2 in \citet{iwasawa16:_implem_fdps} except that the code shown here is written in Fortran. The behavior of the code is very simple: it sets the initial condition by reading particle data from a binary file, and then, it follows the time evolution of the system by integrating Newton's equations of motion of particles with the leap-frog method. The gravitational forces acting on particles are calculated by using FDPS with the tree algorithm in which the gravitational forces from distant particles are approximated as those from a superparticle with multipole moments. For simplicity, we use the center-of-mass approximation. In this case, the gravitational acceleration of particle $i$ is evaluated as the sum of contributions from other particles and superparticles:
\begin{eqnarray}
\frac{\diff^{2}\bm{r}_{i}}{\diff t^{2}} & = \sum_{j=1 \atop (j\neq i)}\frac{Gm_{j}(\bm{r}_{j}-\bm{r}_{i})}{(|\bm{r}_{j}-\bm{r}_{i}|^{2}+\epsilon^{2}_{i})^{3/2}} \nonumber\\
& +\sum_{j=1}\frac{G m'_{j}(\bm{r}'_{j}-\bm{r}_{i})}{(|\bm{r}'_{j}-\bm{r}_{i}|^{2}+\epsilon^{2}_{i})^{3/2}}, \label{eq:nbody_EoM_tree}
\end{eqnarray}
where $\bm{r}_{i}$, $\bm{r}_{j}$, $\bm{r}'_{j}$ are the positions of particle $i$,  particle $j$, and superparticle $j$, respectively. $m_{j}$ and $m'_{j}$ are the masses of particle $j$ and superparticle $j$, respectively. $\epsilon_{i}$ the gravitational softening of particle $i$, $G$ the gravitational constant.

\begin{figure*}[htbp]
\centering
\includegraphics[width=0.95\hsize]{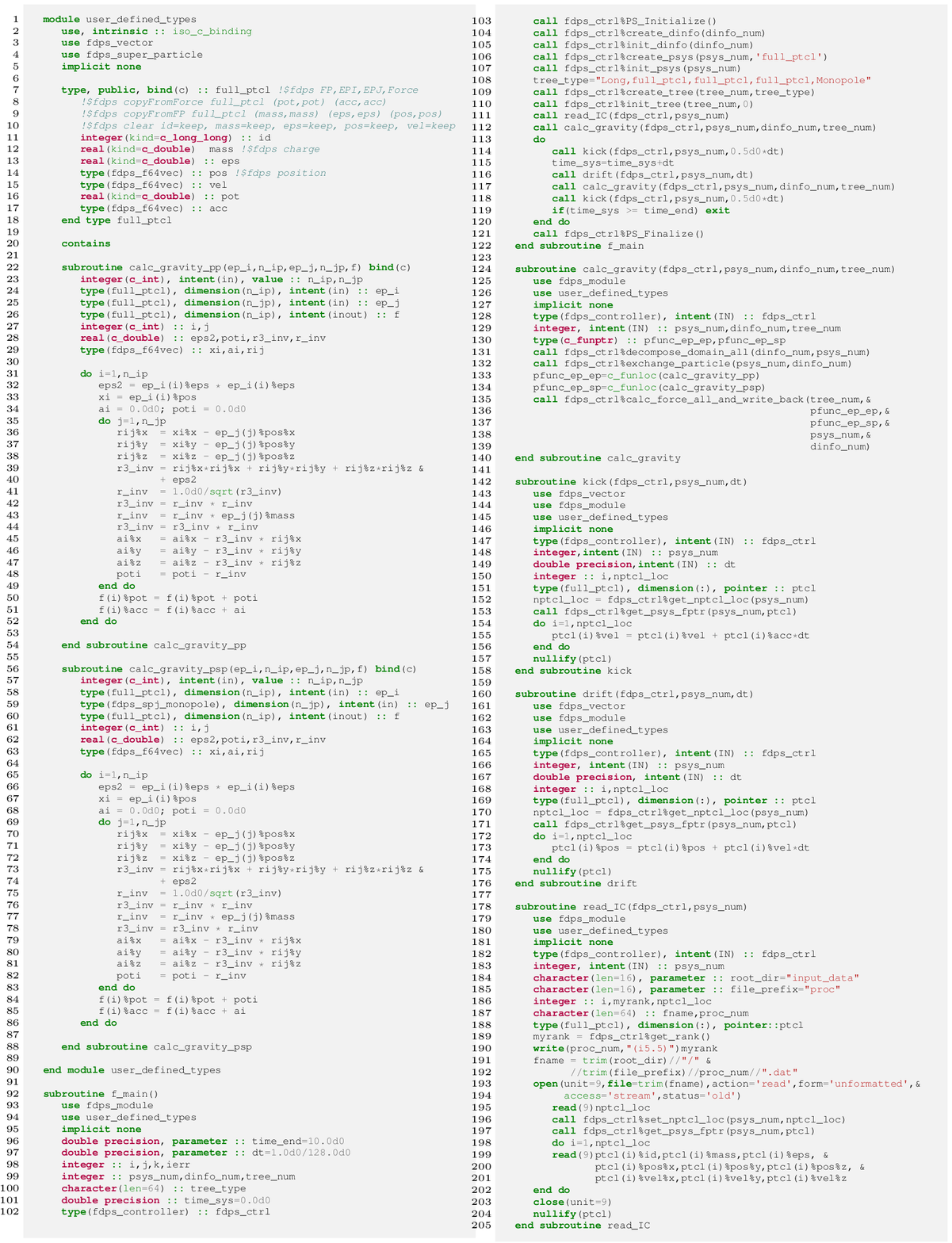}
\caption{A sample code for $N$-body simulation developed by Fortran interface to FDPS.}
\label{fig:src_ftn_smpl}  
\end{figure*}

Figure~\ref{fig:src_ftn_smpl} shows the sample code, which runs either in a single-process execution case or an MPI parallel environment. The code consists of three parts: the definition of the particle (line 7--18), the definition of the interaction functions (line 22--88), and the other parts including numerical integration and I/O (line 92--205). These three parts corresponding to Procedures I,III,IV described at the beginning of this section. In the following, we explain each procedure in detail.

\subsubsection{Defining particle types}
\label{subsubsec:def_ptcl_type}
Users of FDPS must first define particles as derived data types in Fortran (Procedure I). In the sample code, the particle data is defined as \texttt{full\_ptcl} type in lines 7--18. It has the following member variables: \texttt{id} (particle identification number), \texttt{mass} ($m_{i}$), \texttt{eps} ($\epsilon_{i}$), \texttt{pos} ($\bm{r}_{i}$), \texttt{vel} ($\bm{v}_{i}$; the velocity of particle $i$), \texttt{pot} ($\phi_{i}$; the gravitational potential at $\bm{r}_{i}$), \texttt{acc} ($\diff \bm{v}_{i}/\diff t$).

As we will describe in \S~\ref{subsec:ftn_if_impl}, the particle data type must be interoperable with C because Fortran interface uses the C interoperability feature in Fortran 2003 to exchange data between Fortran programs and the core part of FDPS. In order to be interoperable with C, a derived data type must satisfy the following conditions.
\begin{itemize}
\item It has the \texttt{bind(c)} attribute.
\item The data types of all member variables must be interoperable with C.
\end{itemize}
The first condition is satisfied if a derived data type is declared with \texttt{bind(c)} keyword (line 7). As for the second condition, Fortran 2003 or later offers data types corresponding to primitive data types in the C language such as \texttt{int}, \texttt{float}, \texttt{double}, etc.,  through the \texttt{iso\_c\_binding} module. The second condition is fulfilled if we define member variables using these data types. \texttt{integer(kind=c\_long\_long)} is one such example. \texttt{type(fdps\_f64vec)} type used at lines 14, 15, and 17, which is defined in module \texttt{fdps\_vector} and represents a space vector, is also interoperable with C because their member variables are all interoperable with C.

Users of FDPS must write FDPS directives within the definition part of the particles. You can see a variety of Fortran comments which begin with  \texttt{!\$fdps} in the sample code. They are FDPS directives. FDPS directives are used to pass complementary information about particle, e.g. which member variable represents physical quantity FDPS must know, to FDPS. FDPS directives can be classified into the following three types.
\begin{enumerate}[label=(\alph*)]
\item Directive specifying the type of a particle. 
\item Directive specifying which member variable represents which physical quantity.
\item Directive specifying the way of data operation performed in FDPS.  
\end{enumerate}
In the following, we explain each type of directive.

Directive (a) is used to specify which user-defined type a derived data type corresponds to out of \texttt{FP}, \texttt{EPI}, \texttt{EPJ}, and \texttt{Force} described in \S~\ref{subsec:fdps_impl}. A derived data type can serve multiple user-defined types. Directive (a) in this sample code is written at line 7:
\begin{verbatim}
!$fdps FP,EPI,EPJ,Force
\end{verbatim}
With this, \texttt{full\_ptcl} type can be used as any of user-defined types in the sample code.

Directive (b) is used to specify which member variable represents physical quantity required by FDPS. In the present case, we need to specify the mass and the position of a particle (the mass is required to calculate the monopole information of superparticles). Therefore, directives (b) are written as follows (see lines 12 and 14): 
\begin{verbatim}
real(kind=c_double) mass !$fdps charge
type(fdps_f64vec) :: pos !$fdps position
\end{verbatim}
With these, FDPS recognizes that the variables \texttt{mass} and \texttt{pos} represent the mass and the position of a particle. 

In the sample code, three directives of type (c) are written in lines 8--10 as follows.
{\footnotesize
\begin{verbatim}
!$fdps copyFromForce full_ptcl (pot,pot) (acc,acc)
!$fdps copyFromFP full_ptcl (mass,mass) (eps,eps) \
   (pos,pos) 
!$fdps clear id=keep, mass=keep, eps=keep, \
   pos=keep, vel=keep
\end{verbatim}
}
where the symbol \texttt{\textbackslash} is used to represent that the current line continues to the next line because of space limitations and it cannot be used in an actual code (FDPS directive must be written within a line). The first two directives containing keywords \texttt{copyFromForce} and \texttt{copyFromFP} specifies how data copy is performed between different user-defined types. The former specifies the way of data copy from \texttt{Force} type to \texttt{FP} type and the latter specifies that from \texttt{FP} type to \texttt{EPI} type or \texttt{EPJ} type. The directive including keyword \texttt{clear} is used to specify how to initialize \texttt{Force} type before starting the calculations of interactions. 

The detail of FDPS directives can be found in the specification document of Fortran interface in the distributed FDPS package.

\subsubsection{Generating Fortran interface}
\label{subsubsec:gen_ftn_if}
Next step is the generation of Fortran interface programs (Procedure II), which can be simply done by executing the auto-generation script \texttt{gen\_ftn\_if.py} provided by us in the command line as follows.
\begin{verbatim}
./$(FDPS_LOC)/scripts/gen_ftn_if.py sample_code.F90
\end{verbatim}
where \texttt{\$(FDPS\_LOC)} represents the PATH of the top directory of FDPS library. If the auto-generation succeeds, all the files described in \S~\ref{subsubsec:ftn_if_str} are created at the current directory.

\subsubsection{Defining interaction functions}
\label{subsubsec:def_interact_func}
Users of FDPS must define interactions as subroutines in Fortran (Procedure III). In the sample code, two subroutines are defined --- one for the gravitational interaction between particles (lines 22--54) and the other for the interaction between particles and superparticles (lines 56--88).

The interaction functions must also be interoperable with C because their function pointers are passed to the core part of FDPS using the C interoperability feature in Fortran 2003. To be interoperable with C, subroutines satisfies the following conditions.
\begin{itemize}
\item It has the \texttt{bind(c)} attribute.
\item The data types of all variables used in subroutines must be interoperable with C.
\end{itemize}
These conditions are the same as those for derived data types which are interoperable with C (see \S~\ref{subsubsec:def_ptcl_type}). Hence, we can clear these requirements in the same way: first add the \texttt{bind(c)} keyword after the argument list of the subroutine (see lines 22 and 56), and define variables using data types which are interoperable with C.

The interaction functions must have the following arguments. From the beginning, the array of $i$ particles (\texttt{ep\_i}), the number of $i$ particles (\texttt{n\_ip}), the array of $j$ particles or superparticles (\texttt{ep\_j}), the number of $j$ particles or superparticles (\texttt{n\_jp}), and the array of variables that stores the calculated interaction on $i$ particles (\texttt{f}). Due to the specification of the core part of FDPS, the numbers of $j$ particles and superparticles must be pass-by-value arguments. Hence, the \texttt{value} keyword is needed in the type declaration statements for these arguments (see lines 21 and 57).

FDPS does not take care about the optimizations of the interaction functions. So, users of FDPS must optimize the interaction functions by themselves in order to achieve high efficiency. For simplicity, the interaction functions in this sample code are implemented without any optimization techniques. Some typical optimization techniques are presented in \S~\ref{sec:performance}.

\subsubsection{Developing the main part of a user code}
Users of FDPS must implement the main routine of a user code within a subroutine named \texttt{f\_main()} by the reason explained in \S~\ref{subsec:ftn_if_impl} (Procedure IV). \texttt{f\_main()} of the sample code is implemented in lines 92-122 and it consists of the following steps.
\begin{enumerate}
\item Initialize FDPS (line 103)
\item Create and initialize instances of FDPS classes (lines 104-110).
\item Read particle data from a file (line 111).
\item Calculate the gravitational forces on all the particles at the initial time (line 112).
\item Integrate the orbits of all the particles with the leap-frog method (lines 113-120)
\item Finalize FDPS (line 121).
\end{enumerate}
In the following, we first explain the variable declaration part of \texttt{f\_main()}. Then, we explain each step in detail.

In the Fortran interface, FDPS API is provided as type-bound procedures of  Fortran 2003 class \texttt{fdps\_controller} defined in module \texttt{fdps\_module}. This module is defined in \texttt{FDPS\_module.F90}, one of the source programs of Fortran interface. In order to use FDPS API, we first need to make this module accessible from \texttt{f\_main()}, and then we should create an instance of this class. These things are done at lines 93 and 102, respectively. Thus, in this sample code, FDPS API is used by calling type-bound procedures of class instance \texttt{fdps\_ctrl}. 

In step 1, API \texttt{ps\_initialize} is called. This procedure initializes MPI and OpenMP libraries if they are used. If not, it does nothing.

In step 2, we create and initialize three instances of FDPS classes \texttt{ParticleSystem}, \texttt{DomainInfo}, and \texttt{TreeForForce} by calling type-bound procedures whose names contain `\texttt{create}' or `\texttt{init}', which, as the names suggest, create and initialize instances, respectively. Each of these procedures takes an integer argument (\texttt{psys\_num}, \texttt{dinfo\_num}, and \texttt{tree\_num} in the sample code) because all instances of FDPS classes are identified by descriptors in integer variables in the Fortran interface. API \texttt{create\_psys} receives the name of a derived data type representing \texttt{FP} as the second argument, and it creates an instance of \texttt{ParticleSystem<Tfp>}, where \texttt{Tfp} is the name of the C++ class corresponding to the derived data type having the specified name. In the sample code, \texttt{full\_ptcl} is specified. Note that this API accepts only the names of derived data types qualified by FDPS directive of type (a) explained in \S~\ref{subsubsec:def_ptcl_type}. API \texttt{create\_tree} receives the type of an instance of class \texttt{TreeForForce} as the second argument. The type information must be given by a single string of characters that consists of five strings of characters delimited by commas. The first field represents the type of force (long-range or short-range), which is followed by the three names of derived data types corresponding to \texttt{Force}, \texttt{EPI}, and \texttt{EPJ} types, and the final field shows the subtype of tree for long/short-range force. In this example, we calculate the gravitational forces as the long-range force and use the monopole approximation. In addition, \texttt{full\_ptcl} type is used as \texttt{Force}, \texttt{EPI}, \texttt{EPJ} types. Therefore, \texttt{"Long,full\_ptcl,full\_ptcl,full\_ptcl,Monopole"} is specified.

In step 3, we read particle data from a binary file, by using subroutine \texttt{read\_IC} defined in lines 178--205. In the framework of the Fortran interface, all the particle data is stored in a C++ program, which is one of Fortran interface programs (see \S~\ref{subsec:fdps_impl}). In order to setup the particle data from a Fortran program, we should take the following steps.
\begin{enumerate}[label=(\roman*)]
\item Create and initialize an instance of \texttt{ParticleSystem} class (this is already done at step 2).
\item Allocate memory for this instance. In the sample code, this is done by calling API \texttt{set\_nptcl\_loc}. This API allocates a memory for the instance enough to store an array of the particles of size specified by the second argument (i.e. \texttt{nptcl\_loc} in this sample).
\item Get the pointer to this instance. This is done by calling API \texttt{get\_psys\_fptr}. Because this API returns the pointer corresponding to the instance specified by its first argument (i.e., \texttt{psys\_num} in this sample), we have to prepare the pointer of the same type. In the present case, the pointer for an array of \texttt{full\_type} type should be prepared. After the pointer is set, we can use this pointer like an array.
\item Set the particle data using the pointer. This is done in lines 199--201.
\end{enumerate}

In step 4, the interaction calculation is performed through the subroutine \texttt{calc\_gravity}, which is defined in lines 124--140. This subroutine consists of the following three operations.
\begin{enumerate}[label=(\roman*)]
\item Perform domain decomposition. This is done by calling API \texttt{domain\_decomposition} (line 131). The first argument of this API is an integer variable corresponding to an instance of \texttt{DomainInfo} class. The second one is an integer variable corresponding to an instance of \texttt{ParticleSystem} class, which is needed because the positional information of the particles is required to perform domain decomposition.  
\item Perform particle exchange. This is done by calling API \texttt{exchange\_particle} (line 132). This API takes an integer variable corresponding to an instance of \texttt{DomainInfo} class as the first argument. The second argument is an integer variable corresponding to an instance of \texttt{ParticleSystem} class. This is required because the information of subdomains is necessary to perform particle exchange.
\item Perform the calculation of interactions. This is done by calling API \texttt{call\_force\_all\_and\_write\_back} (line 135). The number of the arguments of this API depends on the type of force. In the present case, it takes five arguments because of long-range interaction. The definitions of the first, fourth, fifth arguments should be obvious, and hence we do not explain them here. The second and third arguments are the function pointers for the subroutines that calculate particle-particle interaction and particle-superparticle interaction. The function pointers must be obtained by the intrinsic function \texttt{c\_funloc}, and be stored in variables of \texttt{c\_funptr} type. In the sample code, \texttt{calc\_gravity\_pp} and \texttt{calc\_gravity\_psp} are used for the calculation of interactions.
\end{enumerate}

In step 5, the time integration is performed through the subroutines \texttt{kick} and \texttt{drift}, which are defined, respectively, in lines 142--158 and 160--176. Both subroutines have the same structure: first get the pointer to the particle data stored in the C++ side of Fortran interface programs by using API \texttt{get\_psys\_fptr}, and then, update the particle data using this pointer.

In step 6, API \texttt{ps\_finalize} is called. It calls the \texttt{MPI\_Finalize} function if MPI is used.

In this section, we have described how an application program can be written by using Fortran interface and have shown that researchers can develop particle-based simulation codes by writing Fortran codes only. Thus, the learning of the C++ programing language is totally unnecessary to use FDPS. 

\subsection{Implementation details of Fortran interface}
\label{subsec:ftn_if_impl}
In this section, we describe the implementation details of Fortran interface. In \S~\ref{subsubsec:ftn_if_gen}, we describe why we need a generation of Fortran interface. In \S~\ref{subsubsec:ftn_if_str}, we describe the file structure of Fortran interface programs and the role of each file.

\subsubsection{Necessity of generation of Fortran interface}
\label{subsubsec:ftn_if_gen}
One difficulty in developing a Fortran interface to FDPS is that the template feature in the C++ programing language cannot be used directly from Fortran (Fortran is not designed to interoperate with the C++ programing language). As described earlier, the template feature is essential for FDPS to support arbitrary types of particle and interaction. Hence, we need to workaround this problem in some way in order to make a Fortran interface support arbitrary types of particle and interaction.

In this study, we solve this problem by making use of (i) the fact that Fortran can interoperate with the C programing language through the use of  \texttt{iso\_c\_binding} module introduced in Fortran 2003, (ii) the fact that C++ functions can be called from C programs by using \texttt{extern "C"} modifier, and (iii) auto generation of source codes. The outline of our solution is as follows. Owing to (i) and (ii), we can make Fortran interoperate with C++ via C. Therefore, it is also possible to use a FDPS library (a set of functions to manipulate FDPS) made for a \textit{specific} particle class from a Fortran code through a C interface to the library. However, what we want to realize is to enable researchers to use a FDPS library for an \textit{arbitrary} particle class from a Fortran code. The only way to realize this would be to generate a FDPS library for a given particle class and use it from a Fortran code. Another requirement in designing a Fortran interface is that researchers must be able to develop codes by Fortran only so that they will not need to invest much time in learning C++. To make this possible, we must generate a FDPS library based on some Fortran data structures representing particles, instead of based on C++ classes. It is natural to use Fortran's derived data type, which is similar to structure in C, to define a particle. Thus, we adopted the following solution: first define a particle as a derived data type in Fortran. Then, generate a C++ class corresponding to this derived data type. Finally, generate a FDPS library for this particle class. The generation of required C++ classes and a library is automatically done by a script provided by us. Hence, researchers do not have to care about it.

The one remaining problem is how to generate a C++ class from a given Fortran derived data type. In FDPS, C++ class representing particle must have specific public member functions such as \texttt{getPos} as described in \S~\ref{sec:overview_of_fdps}. Fortran 2003 supports object-oriented programing and offers classes for it, which are derived data types having type-bound procedures and are essentially same as classes in C++. Thus, a simple way to generate a C++ class is to define particle as a class in Fortran 2003 that has public member functions required by FDPS and to convert it to a C++ class. However, we cannot take this approach because Fortran 2003 Standard specifies that derived data types must not have member procedures/functions to be interoperable with C. In order to solve this problem, we introduce FDPS directives. They are used for users of FDPS to pass all information necessary to generate public member functions of the corresponding C++ classes to the auto-generation script described above. For example, the auto-generation script needs to know which member variable of a given Fortran derived data type represents the position of a particle in order to generate a public member function \texttt{getPos} in the corresponding C++ class. This information is passed to the script by adding FDPS directive \texttt{!\$fdps position} to the corresponding member variable of the Fortran derived data type. In other words, our auto-generation script is designed to generate \texttt{getPos} in the corresponding C++ class using a member variable of a given Fortran derived data type indicated by FDPS directive \texttt{!\$fdps position}. In this way, we can pass all of the necessary information to the auto-generation script. In \S~\ref{subsec:ftn_if_usage}, we have given a rough explanation about FDPS directive for simplicity, but actually this kind of things is done.

\begin{figure}[h]
\begin{center}
\includegraphics[width=8cm]{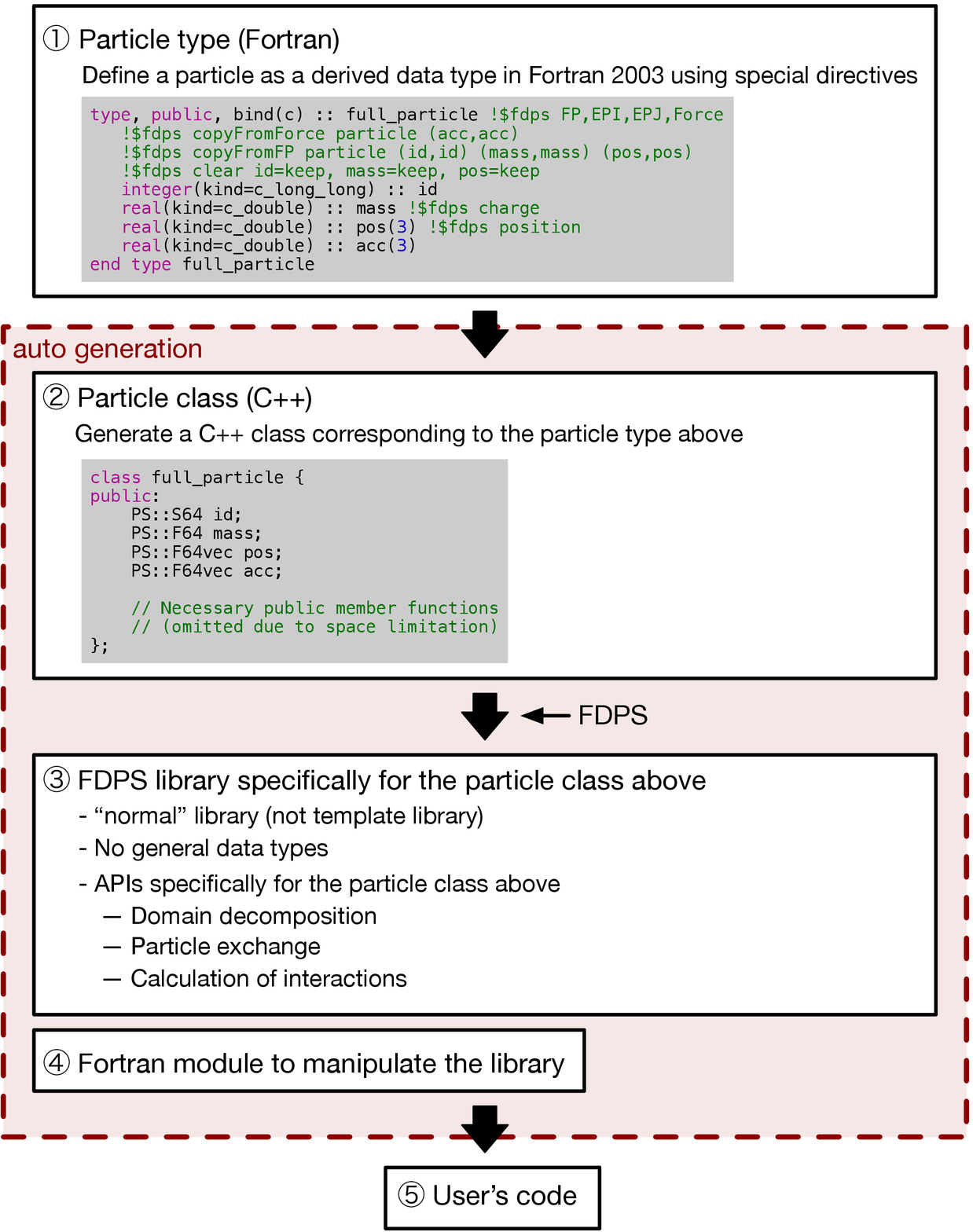}
\end{center}
\caption{A schematic illustration of our solution to enable to use arbitrary type of particle in a Fortran interface.}
\label{fig:our_solution}  
\end{figure}

The generation of a Fortran interface can be summarized schematically as shown in Figure~\ref{fig:our_solution}: Firstly, researchers define particles as derived data types in Fortran with FDPS directives (Step {\large \lower1pt\hbox{\ding{192}}} in Fig.~\ref{fig:our_solution}). Next, the auto-generation script generates C++ classes corresponding to the given Fortran derived data types (Step {\large \lower1pt\hbox{\ding{193}}}). Using the information given by FDPS directives, the script can generate public member functions required by FDPS. Then, the script generates a FDPS library specifically for these C++ classes (Step {\large \lower1pt\hbox{\ding{194}}}). The generated library is not a template library, but a normal library that is callable in C manner. Finally, the script also generates a Fortran module to manipulate the generated library (Step {\large \lower1pt\hbox{\ding{195}}}). Users of FDPS can use FDPS through this module (Step {\large \lower1pt\hbox{\ding{196}}}).

\subsubsection{Structure of Fortran interface programs}
\label{subsubsec:ftn_if_str}
In this section, we explain the file structure of the generated programs and their internal structures.

\begin{figure}[h]
\begin{center}
\includegraphics[width=8cm]{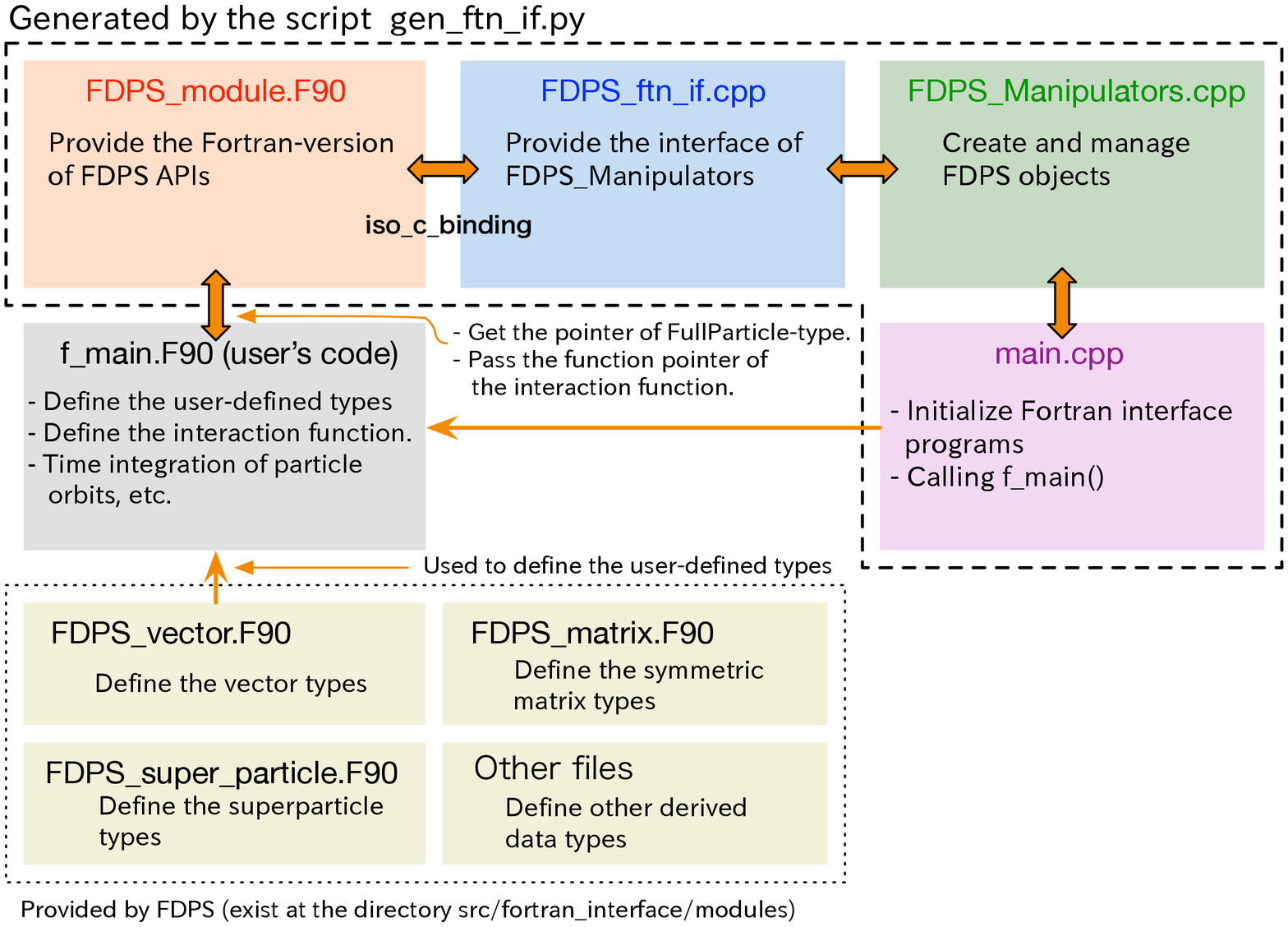}
\end{center}
\caption{File structure of Fortran interface programs and their relations to user’s code}
\label{fig:file_str_ftn_if}  
\end{figure}

Figure~\ref{fig:file_str_ftn_if} is a brief summary of the file structure of the Fortran interface programs and their roles. Four files enclosed by the dashed line (\texttt{FDPS\_module.F90}, \texttt{FDPS\_ftn\_if.cpp}, \texttt{FDPS\_Manipulators.cpp}, \texttt{main.cpp}) are the files to be generated by the script, and the file \texttt{f\_main.F90} corresponds to user programs. In the files enclosed by the dotted line (\texttt{FDPS\_vector.F90}, \texttt{FDPS\_matrix.F90}, \texttt{FDPS\_super\_particle.F90}, etc.), several derived data types are defined, which are needed to define user-defined types and interaction functions in Fortran (see also \S~\ref{subsec:ftn_if_usage}). In the following, we explain the role of each interface program.

At first, we explain the roles of \texttt{FDPS\_Manipulators.cpp} and \texttt{main.cpp}. Because the core part of FDPS is written in C++, all of C++ instances of FDPS classes described in \S~\ref{sec:overview_of_fdps} (i.e. \texttt{DomainInfo}, \texttt{ParticleSystem}, and \texttt{TreeForForce} classes) must be created and be managed in C++ codes. This task is performed by \texttt{FDPS\_Manipulators.cpp}, which corresponds to a library shown in Step {\large \lower1pt\hbox{\ding{194}}} in Fig.~\ref{fig:our_solution}. By the same reason, we must place the main function of the user program in a C++ file. Thus, \texttt{main.cpp} is generated. It calls a Fortran subroutine named \texttt{f\_main()}. Hence, users should prepare a Fortran subroutine \texttt{f\_main()} and must implement all parts of the simulation code inside \texttt{f\_main()}. In the Fortran interface, all of C++ instances created in \texttt{FDPS\_Manipulators.cpp} are assigned to Fortran’s integer variables. Hence, users also need to manage these instances using integer variables.

The file \texttt{FDPS\_ftn\_if.cpp} provides C interfaces to the functions defined in \texttt{FDPS\_Manipulators.cpp}. These C functions can be called from a Fortran program by using the functionalities provided by module \texttt{iso\_c\_binding} in Fortran 2003 as described in \S~\ref{subsubsec:ftn_if_gen}. The file \texttt{FDPS\_module.F90} provides a class in Fortran 2003, named \texttt{fdps\_controller}, for users, which is used to call the C interface functions described above. All of the FDPS API for Fortran are provided as public type-bound procedures of this class. Therefore, in order to use the FDPS API, we first need to create an instance of this class, and then call its type-bound procedures.

As described above, the particle data is stored in \texttt{FDPS\_Manipulators.cpp} as explained above. It is actually an array of C structures representing particles (cf. Step {\large \lower1pt\hbox{\ding{193}}} in Fig.~\ref{fig:our_solution}). Fortran module \texttt{iso\_c\_binding} enables us to access it if a derived data type in Fortran corresponding to this C structure is interoperable with C. The Fortran interface makes use of this functionality, and therefore, we require that all derived data types that will be used in FDPS must be interoperable with C. We also use the functionalities of module \texttt{iso\_c\_binding} to pass interaction functions to \texttt{FDPS\_Manipulators.cpp}. To take this approach, interaction functions must be implemented as subroutines in Fortran and they must be interoperable with C. This is because subroutines are passed in the form of the C addresses  (i.e. function pointers in C), which can be obtained only if the subroutines are interoperable with C. These are the reasons why we require C interoperability in \S~\ref{subsec:ftn_if_usage}.


\section{Performance of application developed by Fortran interface}
\label{sec:performance}
In this section, we present and discuss the performance of an application developed by Fortran interface to FDPS. Here, we mainly focus on the overhead of the FDPS API-call from a Fortran application because all the tasks of the API are processed in its C++ core and it is expected that the times required to process these tasks do not depend on development language. To this end, we compare difference between the performances of Fortran and C++ codes for $N$-body simulations. The performance measurements are performed in different computer systems to check the dependencies of the performance on CPU architectures and compilers. Table~\ref{tbl:computer_systems} lists the computer systems and the compiler information used in the measurements.

\begin{longtable}{l|p{4cm}p{4cm}p{4cm}}
\caption{Computer systems and compiler information}
\label{tbl:computer_systems}
\hline
System name & K Computer & Oakforest-PACS & Cray XC30   \\
\hline\hline
\endfirsthead
System name & K Computer & Oakforest-PACS & Cray XC30   \\
\hline\hline
\endhead
\hline
\endfoot
\hline
\endlastfoot
CPU & Fujitsu SPARC 64 VIIIfx & Intel Xeon Phi 7250 & Intel Xeon CPU E5-2650 v3  \\
Compiler & Fujitsu compilers  & Intel compilers  & Intel compilers  \\
& (ver. 1.2.0 P-id: L30000-15) & (ver. 17.0.4 20170411) & (ver. 16.0.4 20160811) \\
Compile options (C++) & \texttt{-Kfast -Ksimd=2 -Krestp=all} & \texttt{-O3 -ipo -xMIC-AVX512 -no-prec-div} & \texttt{-fast -ipo -xCORE-AVX2 -no-prec-div} \\
Compile options (F) & \texttt{-X 03 -Kfast -Ksimd=2 -Free -fs} & \texttt{-O3 -ipo -xMIC-AVX512 -no-prec-div} & \texttt{-fast -ipo -xCORE-AVX2 -no-prec-div} \\
\end{longtable}

\begin{figure}[h]
\begin{center}
\includegraphics[width=8cm]{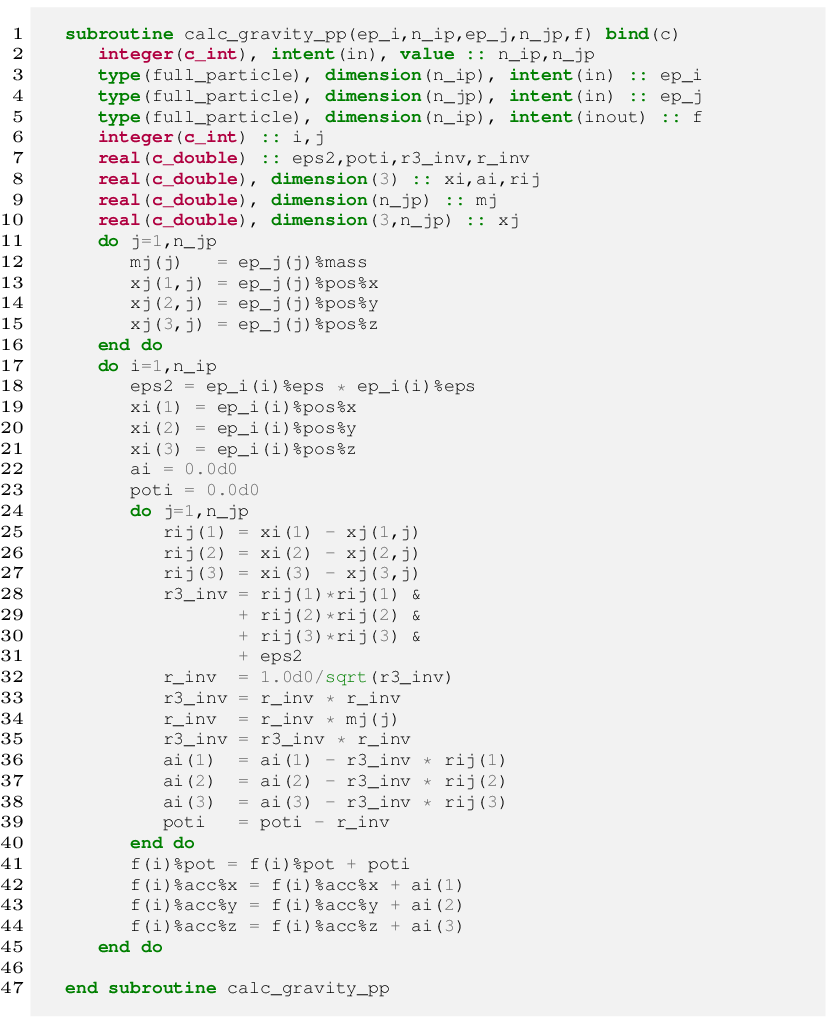}
\end{center}
\caption{Interaction function in the $N$-body simulation code written in Fortran.}
\label{fig:src_ftn_nbody_kernel}  
\end{figure}

\begin{figure}[h]
\begin{center}
\includegraphics[width=8cm]{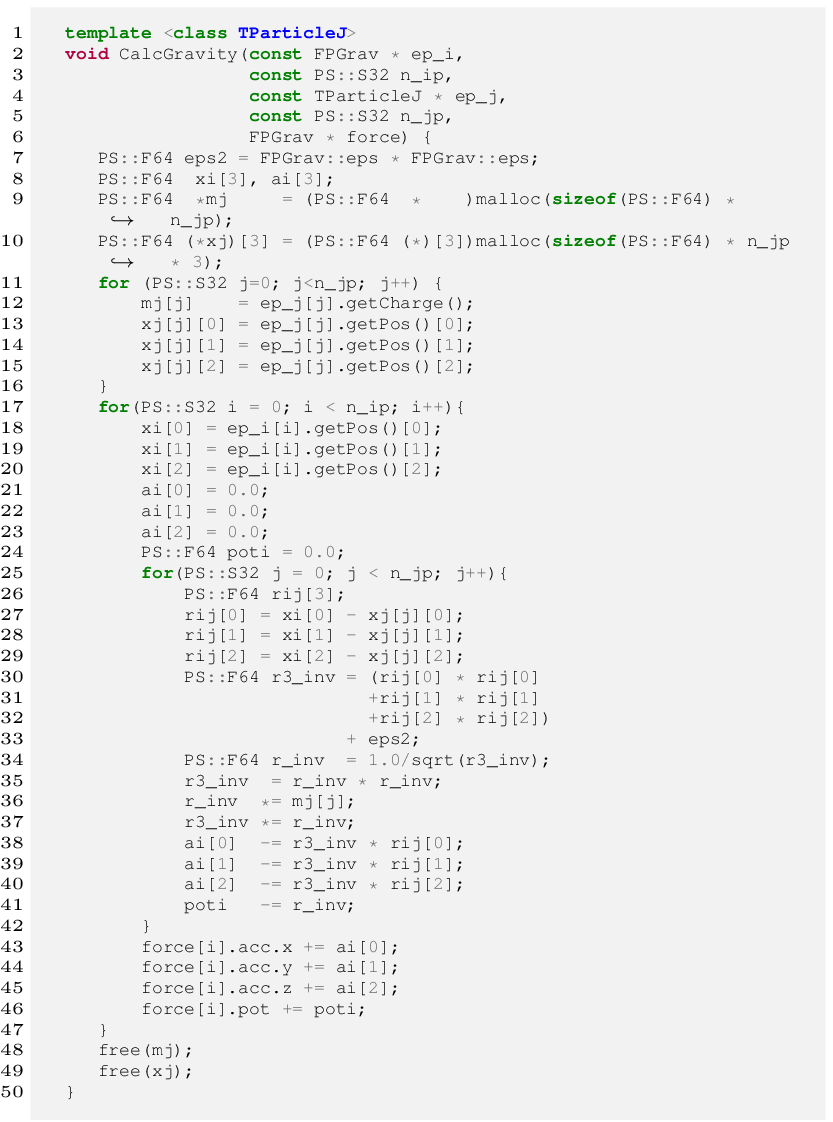}
\end{center}
\caption{Interaction function in the $N$-body simulation code written in C++, where the symbol $\hookrightarrow$ represents that the current line is a continuation line to the previous line.}
\label{fig:src_cpp_nbody_kernel}  
\end{figure}

The Fortran code used in the performance measurement is essentially same as the sample code described in \S~\ref{subsec:ftn_if_usage} (Fig.~\ref{fig:src_ftn_smpl}) except the following two differences. First, we apply standard optimization techniques to interaction functions. Figure~\ref{fig:src_ftn_nbody_kernel} shows a Fortran subroutine for the gravitational interaction between particles. In the subroutine, the information of $j$-particles are stored into local arrays (lines 11-16), and all the calculations in the innermost loop are described using arrays (lines 25-39). These modifications are aiming to facilitate compilers to optimize the code easier. We use almost the same subroutine for the interaction between particles and superparticles. For a fair comparison, in the C++ version of the $N$-body simulation code, we use a function with the same structure, which is shown in Fig.~\ref{fig:src_cpp_nbody_kernel}.

Second, we rewrite all the operations between derived data types by using their member variables explicitly. This is because in the combination of Intel Xeon Phi and intel compilers the executable file which is generated with the highest optimization compile flags does not perform some numerical operations correctly. We believe that this is the problem on the compiler side and this situation will be improved by future upgrade of the compiler. But for now, we recommend users of Fortran interface to follow this prescription.

\begin{figure*}[h]
\begin{center}
\includegraphics[width=0.95\linewidth]{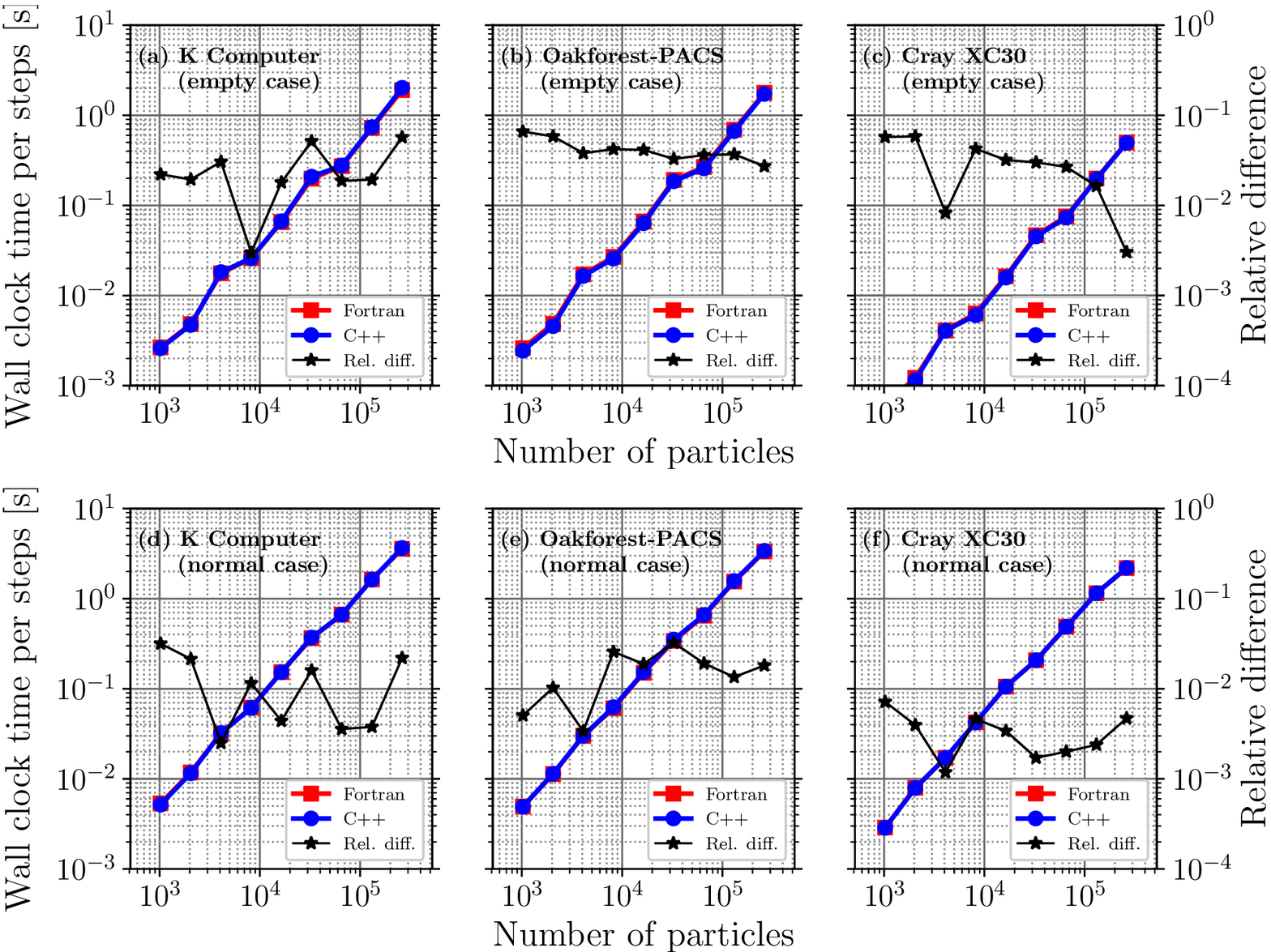}
\end{center}
\caption{Wall-clock time per steps for different numbers of particles for each computer system (upper row: the case when the interaction functions are empty, lower row: the case when the interaction functions shown in Figs.~\ref{fig:src_ftn_nbody_kernel} and \ref{fig:src_cpp_nbody_kernel} are used; left: K-Computer, middle: Oakforest-PACS, right: Cray XC30). Relative difference of the wall-clock times between Fortran code and C++ code is also shown by stars.}
\label{fig:nbody_comp}  

\end{figure*}

The both (Fortran and C++) codes solve the cold collapsed problem, which is one of the standard test problems in $N$-body simulation codes. The most of the calculation times are spent on the part of the interaction calculation and the overhead due to Fortran interface may be hidden by this part. In order to clarify the cost of the overhead, we measure the performances for the following two cases: One is the case when the interaction functions are empty (empty cases) and the other is the case when the interaction functions shown in Figs.~\ref{fig:src_ftn_nbody_kernel} and \ref{fig:src_cpp_nbody_kernel} are used (normal cases). Both codes are executed with a single core. In this case, the APIs for domain decomposition and particle exchange do nothing and just increase the overhead. The upper panels of Fig.~\ref{fig:nbody_comp} show the wall-clock time per step for different number of particles per processes for each computer system for the former case, while the lower panels show those for the latter case. As shown in the upper panels, the calculation times in both codes are almost the same for all computer systems we used. Relative differences of the wall-clock times between both codes are $\approx 5\%$. Thus, the overhead due to Fortran interface is sufficiently small. The overall performances also show a similar behavior as shown in the lower panels of Fig.~\ref{fig:nbody_comp}. Compared to the case of empty interaction function, the relative differences become small in each computer system. Thus, a code written by Fortran interface shows a performance nearly identical to that written by C++.

\section{Example of practical application}
\label{sec:example}
%
In order to demonstrate its ability, we perform a large-scale global planetary ring simulation using a code developed by Fortran interface. Ring particles around a planet interact with each other through mutual gravity and inelastic collision. In the code, we calculate the gravity using the Tree method with an opening angle criterion of $\theta=0.5$. The inelastic collision is modeled by the soft sphere model following \citet{salo95:simulat_of_dense_planetary} and \citet{michikoshi17a:_simulat_smallest}. The initial condition is taken from \citet{michikoshi17a:_simulat_smallest}, who investigated the dynamics of two narrow rings around Centaur Chariklo with changing the radius and mass of ring particles. Among their models, we consider the case of $r_{p}=5\;\mathrm{m}$ and $\rho_{p}/\rho_{C}=0.5$, where $r_{p}$ and $\rho_{p}$ are the radius and density of ring particles, respectively. $\rho_{C}=1.0\;\mathrm{g\;cm^{-3}}$ is the density of Chariklo. For simplicity, we consider the inner ring only, which is located at a distance of $a=390.6\; \mathrm{km}$ from the center of Chariklo. The radial width and optical depth of the ring are $6.7\;\mathrm{km}$ and $0.38$, respectively.  We model the ring by using $79557408$ particles. With these parameters, it is expected that the self-gravitational wakes form in the ring in a dynamical time (\cite{toomre64:on_the_grav_instab_}). The simulation is performed with using $1088$ MPI processes and $4$ OpenMP threads on Oakforest-PACS.

\begin{figure*}[htbp]
\begin{center}
\includegraphics[width=0.95\hsize]{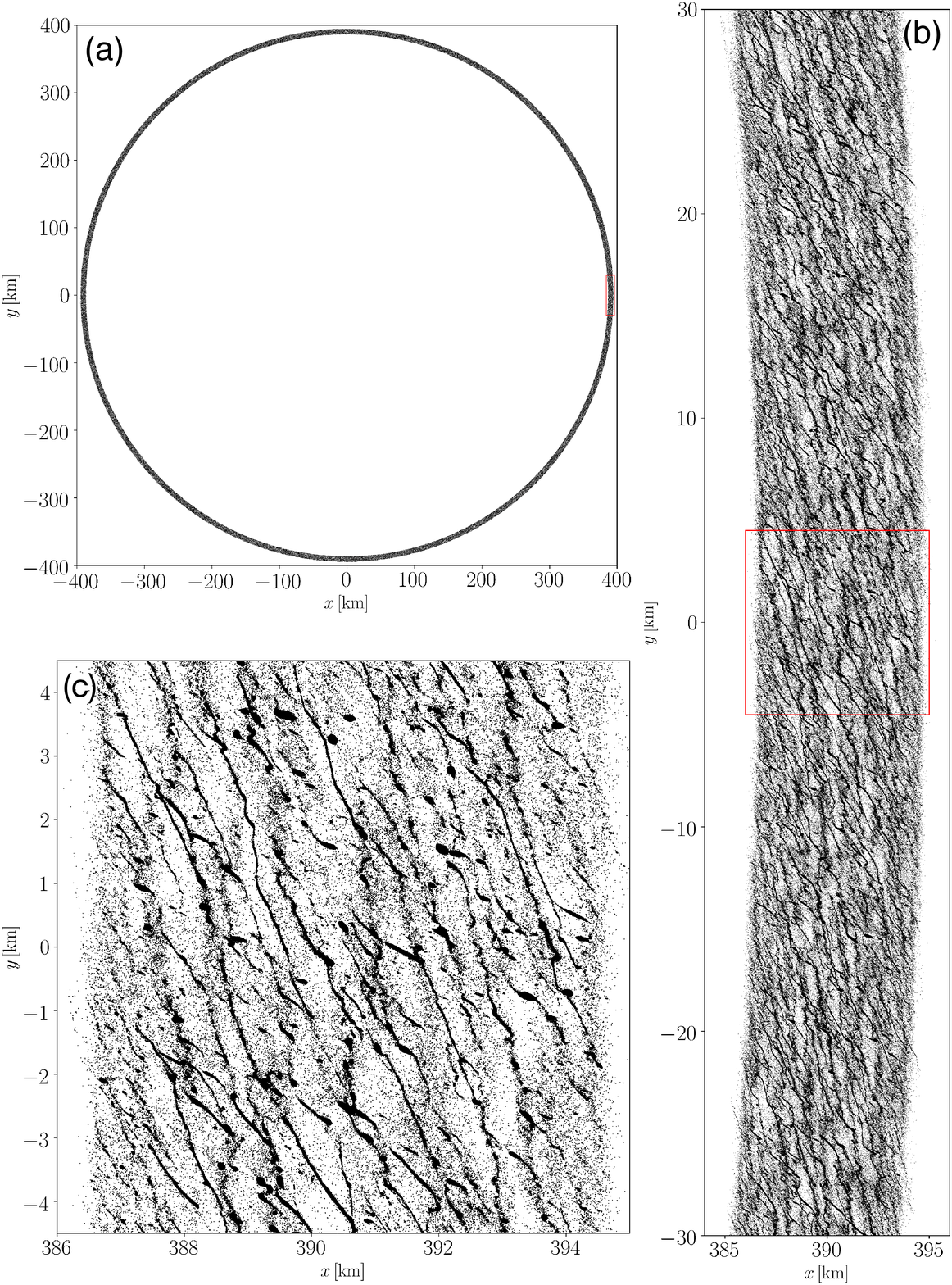}  
\end{center}
\caption{The distribution of ring particles at $t=10t_{\mathrm{Kep}}$ in the planetary ring simulation. Panels (b) and (c) are enlarged views of regions surrounded by the red rectangles in panels (a) and (b), respectively. In panel (a), we plot only a tenth of particles in order to keep the file size of the figure down.}
\label{fig:planetary_ring}  
\end{figure*}

Figure~\ref{fig:planetary_ring} show the distribution of ring particles at $t=10t_{\mathrm{Kep}}$, where $t_{\mathrm{Kep}}$ is the orbital period at the distance of $a$. The particle distribution closely resembles that of \citet{michikoshi17a:_simulat_smallest} as expected (see their Fig.~1). 

Based on our experience, a great deal of time was not needed to complete this application. Thus, by using Fortran interface, researchers can concentrate on their study without spending a lot of time to develop simulation codes.

\section{Summary}
\label{sec:summary}
In this paper, we have presented the basic design and the implementation of the  Fortran interface layer of FDPS, which is newly introduced in version 3.0. This layer provides API for Fortran and supports arbitrary types of particles and interactions as with the C++ core part of FDPS. This is realized by means of both a feature introduced in Fortran 2003 and auto-generation of interface programs. The Fortran interface also support almost all the standard features of FDPS. By using the Fortran interface, researchers can develop massively-parallel particle simulation codes in Fortran.

We also have presented the performance of an application developed by Fortran interface. By comparing with the performance of a C++ version of the application, we have shown that the cost of the overhead due to Fortran interface is sufficiently small and the overall performances of both codes are very similar to each other.

\bigskip
We are grateful to M.~Tsubouchi, Y.~Wakamatsu, and Y.~Yamaguchi for their help in managing the FDPS development team. This research used computational resources of the K computer provided by the RIKEN Advanced Institute for Computational Science through the HPCI System Research project (Project ID:ra000008). Part of the research covered in this paper research was funded by MEXT’s program for the Development and Improvement for the Next Generation Ultra High-Speed Computer System, under its Subsidies for Operating the Specific Advanced Large Research Facilities. Numerical computations were in part carried out on Cray XC30 at Center for Computational Astrophysics, National Astronomical Observatory of Japan, and on Oakforest-PACS at Supercomputing Division, Information Technology Center, the University of Tokyo. L.W. in this work is supported by the funding from Alexander von Humboldt Foundation.


\end{document}